\documentclass[onecolumn,pra]{revtex4}
\usepackage{bm}
\usepackage{epsfig}
\usepackage{amsmath}
\begin{document}
\title{Long Range Electron Transfer Reactions: An Analytically Solvable Model}
\author{Aniruddha Chakraborty \\
School of Basic Sciences, Indian Institute of Technology Mandi,\\
Mandi, Himachal Pradesh, 175001, India}
\date{\today }
\begin{abstract}
\noindent  We propose an analytical method for understanding the problem of long range electron transfer reaction in solution, modeled by a particle undergoing diffusive motion under the influence of many potentials which are involved (donor - bridge - acceptor) in the process. The coupling between these potentials are assumed to be represented by Dirac Delta functions. The diffusive motion in this paper is represented by the Smoluchowskii equation. Our solution requires the  knowledge of the Laplace transform of the Green's function for the motion in all the uncoupled potentials. 
\end{abstract}
\maketitle
Understanding of electron transfer processes in condensed phase is very important in chemistry, physics, and biological sciences, for the experimentalists as well as theoreticians \cite{Marcus,MarcusRev,Zusman1,Zusman2,Zusman3,Wolynes1,Wolynes2,Hynes,Barbara1,Barbara2,Myers,Ivanov,Jortner,Cukier,Sumi,Maroncelli,Dutton1,Dutton2,Bagchi,AniJCP}. A large amount of research in this area has been dedicated in the understanding of the behavior of electron transfer reactions exhibited by donor-acceptor pairs in solutions. Long range electron transfer in condensed phase may occur over long distances (upto several tens of Angstroms), plays a key role in many physical, chemical and bilogical procsses \cite{Davidson}. In the following we propose a simple analytical method for understanding the problem of long range electron transfer reaction in solution, modeled by a particle undergoing diffusive motion under the influence of many (donor - bridge - acceptor) potentials explicitly.  A molecule (donor - bridge - acceptor) immersed in a polar solvent can be put on an electronically excited potential (represents the free energy of the donor surface) by the absorption of radiation. The molecule executes a walk on that potential, which may be taken to be random as it is immersed in the polar solvent. As the molecule moves it may undergo non-radiative decay from certain regions of that potential to another potential (represents the free energy of one, among many which constitutes bridge potentials). So the problem is to calculate the probability that the molecule will still be on the electronically excited donor potential after a finite time $t$.
We denote  the probability that the molecule would survive on the donor potential by $P_{d}(x,t)$. We also use $P_{a}(x,t)$ to denote the probability that the molecule would be found in the acceptor potential and $P_{i}(x,t)$ to denote the probability that the molecule would be found on the $i$-th bridge potential. It is very usual to assume the motion on all the potentials to be one dimensional and diffusive, the relevent coordinate being denoted by $x$. It is also common to assume that the motion on all the potential energy surface is overdamped. Thus all the probability $P_{d}(x,t)$, $P_{i}(x,t)$ and $P_{a}(x,t)$ may be found at $x$ at the time $t$ obeys a modified Smoluchowskii equation.
\begin{eqnarray}
\frac{\partial P_d(x,t)}{\partial t} = {\cal L}_d P_d(x,t)+k_r P_d(x,t)- k_0 S(x) P_{1}(x,t) \\ \nonumber
\frac{\partial P_1(x,t)}{\partial t} = {\cal L}_1 P_1(x,t)+k_r P_1(x,t)- k_0 S(x) P_{2}(x,t)- k_0 S(x) P_{d}(x,t) \\ \nonumber
\frac{\partial P_2(x,t)}{\partial t} = {\cal L}_2 P_2(x,t)+k_r P_2(x,t)- k_0 S(x) P_{3}(x,t) - k_0 S(x) P_{1}(x,t) \\ \nonumber
..................................................................................................................  \\ \nonumber
...................................................................................................................  \\ \nonumber
\frac{\partial P_a(x,t)}{\partial t} ={\cal L}_a P_a(x,t)+k_r P_a(x,t) - k_0 S(x) P_{N}(x,t). \nonumber
\end{eqnarray}
In the above 
\begin{equation}
{\cal L}_i= D \left[ \frac{\partial^2}{\partial x^2}+ \beta \frac{\partial}{\partial x} \frac{dV_i(x)}{dx}\right].
\end{equation}
$V_i(x)$ is the potential causing the drift of the particle, $S(x)$ is a position dependent sink function, $k_r$ is the rate of radiative decay and $k_0$ is the rate of electron transfer. We have taken $k_r$ to be independent of position. $D$ is the diffusion coefficient. 
Before we excite, the molecule is in the ground state, and as the solvent is at a finite temperature, its distribution over the coordinate $x$ is random. From this it undergoes Franck-Condon excitation to the excited state potential (donor). So, $x_0$ the initial position of the particle, on the excited state potential is random. We assume it to be given by the probability density $P^{0}_{1} (x_0)$. In the following we provide a general procedure for finding the exact analytical solution of Eq. (1). The Laplace transform ${\cal P}_i(x,s)=\int_{0}^{\infty} P_i(x,t)e^{-st} dt$ obeys
\begin{eqnarray}
[s-{\cal L}_d+k_r] {\cal P}_d(x,s)+k_0 S(x) {\cal P}_1 (x,s) = P^0_d(x_0) \\ \nonumber
[s-{\cal L}_1+k_r] {\cal P}_1(x,s)+k_0 S(x) {\cal P}_2 (x,s)+k_0 S(x) {\cal P}_d (x,s) = 0, \\ \nonumber
[s-{\cal L}_2+k_r] {\cal P}_2(x,s)+k_0 S(x) {\cal P}_3 (x,s)+k_0 S(x) {\cal P}_1 (x,s) = 0, \\ \nonumber
.............................................................................................\\ \nonumber
..............................................................................................\\ \nonumber
[s-{\cal L}_a+k_r] {\cal P}_a(x,s)+k_0 S(x) {\cal P}_N (x,s) = 0,
\end{eqnarray}
where $P^0_d(x_0)=P_d(x,0)$ is the initial distribution at the electronically excited  state (donor potential), $P_i(x,0)=0$ and $P_a(x,0)=0$ is the initial distribution at the acceptor potential. 
\begin{equation}
\left(
\begin{array}{c}
{\cal P}_d (x,s)\\
{\cal P}_1 (x,s)\\
{\cal P}_2 (x,s)\\
................\\
............... \\
{\cal P}_a (x,s)
\end{array}
\right) = \left(
\begin{array}{cccc}
s - {\cal L}_d +k_r & k_0 S(x) & 0 &...... \\
k_0 S(x) & s-{\cal L}_1+k_r &  k_0 S(x) & 0  ....\\
0 & k_0 S(x) & s-{\cal L}_2+k_r &  k_0 S(x) \\
0   & 0 & k_0 S(x) & ... \\
...   &....& .... & ... \\
 ... & ...  & ...  & ...  
\end{array}
\right)^{-1}
\left(
\begin{array}{c}
P^0_d(x_0) \\
0 \\
0\\
...\\
...\\
0
\end{array}
\right)  ,
\end{equation}
The solution of the above equation can be written as 
\begin{equation}
{\cal P}_{d \rightarrow a}(x,s)=\int_{-\infty}^{\infty} dx_0 G^{0}_{d \rightarrow a}(x,s;x_0)P^0_d(x_0),
\end{equation}
where $G_{d \rightarrow a}(x,s;x_0)$ is the corresponding Green's function. In the following we will derive an analytical expression for the Greens function. We start with the simplest version of the problem {\it i.e.} given below
\begin{equation}
 \left(
\begin{array}{c}
{\cal P}_d (x,s) \\
{\cal P}_1 (x,s)
\end{array}
\right) = \left(
\begin{array}{cc}
s-{\cal L}_d+k_r & k_0 S(x) \\
k_0 S(x) & s-{\cal L}_1+k_r
\end{array}
\right)^{-1}
\left(
\begin{array}{c}
P^0_d(x) \\
0
\end{array}
\right)  ,
\end{equation}
Using the partition technique \cite{Lowdin}, solution of this equation can be written as 
\begin{equation}
{\cal P}_{d \rightarrow 1}(x,s)=\int_{-\infty}^{\infty} dx_0 G^{0}_{d \rightarrow 1}(x,s;x_0)P^0_d(x_0),
\end{equation}
where $G(x,s;x_0)$ is the Green's function defined by the following equation
\begin{equation}
G^{0}_{d \rightarrow 1}(x,s;x_0)=\left < x \left|[s-{\cal L}_d+ k_r + {k_0}^2 S[s-{\cal L}_1+k_r]^{-1}S]^{-1}\right| x_0 \right>
\end{equation}
The above equation is true for any general sink $S$. But this expressions simplify considerably if $S$ is a Dirac Delta function located at $x_1$. In the operator notation $S$ can be written as $S= \left| x_1 \left> \right < x_1 \right |$. So
\begin{equation}
G^{0}_{d \rightarrow 1}(x,s;x_0)=\left < x \left|[s-{\cal L}_d+ k_r + {k_0}^2 G^{0}_1(x_1,s;x_1) S ]^{-1}\right| x_0 \right>,
\end{equation}
where
\begin{equation}
G^{0}_1(x,s;x_0)=\left < x \left|[s-{\cal L}_1+ k_r ]^{-1}\right| x_0 \right>
\end{equation}
and corresponds to propagation of the particle starting from $x_0$ on the first bridge potential in the absence of any coupling. Now we use the operator identity
\begin{equation}
[s-{\cal L}_d + k_r + {k_0}^2 G^{0}_1(x_1,s;x_1) S]^{-1}=[s-{\cal L}_d+ k_r]^{-1}-[s-{\cal L}_d+ K_r]^{-1}{k_0}^2 G^{0}_1(x_1,s;x_1) S [s-{\cal L}_d + k_r - {k_0}^2 G^{0}_1(x_1,s;x_1) S]^{-1}
\end{equation}
Inserting the resolution of identity $I=\int_{-\infty}^{\infty} dy \left|y \left> \right < y \right|$ in the second term of the above equation, we arrive at an equation which is very similar to Lippman-Schwinger equation.
\begin{equation}
G^{0}_{d \rightarrow 1}(x,s;x_0)=G^0_d(x,s;x_0) - {k_0}^2 G^0_d(x,s;x_1)G^0_1(x_1,s;x_1)G^{0}_{d \rightarrow 1}(x_1,s;x_0).
\end{equation}
where $G^0_d(x,s;x_0)=\left < x \left|[s-{\cal L}_d+k_r]^{-1}\right| x_0 \right>$ corresponds to the propagation of the particle on donor potential put initially at $x_0$, in the absence of any coupling, it is actually the Laplace Transform of $G^0_d(x,t;x_0)$, which is the probability that a particle starting at $x_0$ can be found at $x$ at time $t$. We now put $x=x_1$ in the above equation and solve for $G^0_{d \rightarrow 1}(x_1,s;x_0)$ to get
\begin{equation}
G^0_{d \rightarrow 1}(x,s;x_0)=\frac{G^0_d(x,s;x_0)}{1+ {k_0}^2 G^0_d(x_1,s;x_1)G^0_1(x_1,s;x_1)}.
\end{equation}
This when substitued back into Eq. (12) gives
\begin{equation}
G^{0}_{d \rightarrow 1}(x,s;x_0)=G^0_d(x,s;x_0) - \frac{{k_0}^2 G^0_d(x,s;x_1)G^0_1(x_1,s;x_1)G^0_d(x_1,s;x_0)}{1+{k_0}^2 G^0_d(x_1,s;x_1)G^0_1(x_1,s;x_1)},
\end{equation}
where $G^0_d(x,s;x_0)=\left < x \left|[s-{\cal L}_d+k_r]^{-1}\right| x_0 \right>$ corresponds to the propagation of the particle on reactant potential put initially at $x_0$, in the absence of any coupling. So if we know the analytical form of both $G^0_d(x,s;x_0)$ and $G^0_1(x,s;x_0)$, we can derive an analytical expression for $G^{0}_{d \rightarrow 1}(x,s;x_0)$. Now we consider the following case, the donor potential and two bridge potentials, so that Eq. (6) will be modified to the following one
\begin{equation}
 \left(
\begin{array}{c}
{\cal P}_d (x,s) \\
{\cal P}_1 (x,s) \\
{\cal P}_2 (x,s) 
\end{array}
\right) = \left(
\begin{array}{ccc}
s-{\cal L}_d+k_r & k_0 S(x) & 0 \\
k_0 S(x) & s-{\cal L}_1+k_r & k_0 S(x)\\
0 &  k_0 S(x) & s-{\cal L}_2+k_r 
\end{array}
\right)^{-1}
\left(
\begin{array}{c}
P^0_d(x) \\
0\\
0
\end{array}
\right)  ,
\end{equation}
Using the partition technique \cite{Lowdin}, solution of this equation can be written as 
\begin{equation}
{\cal P}_{d \rightarrow 2}(x,s)=\int_{-\infty}^{\infty} dx_0 G^{0}_{d \rightarrow 2}(x,s;x_0)P^0_d(x_0),
\end{equation}
where $G^{0}_{d \rightarrow 2}(x,s;x_0)$ can be derived from $G^{0}_{d \rightarrow 1}(x,s;x_0)$ and $G^0_2(x_1,s;x_1)$ using the same method as we have used in deriving Eq.(24), with the assumtion that the second bridge potential in coupled to first one via Dirac delta function at $x_2$.
\begin{equation}
G^{0}_{d \rightarrow 2}(x,s;x_0)=G^0_{d \rightarrow 1}(x,s;x_0) - \frac{{k_0}^2 G^0_{d \rightarrow 1}(x,s;x_2)G^0_2(x_2,s;x_2)G^0_{d \rightarrow 1}(x_2,s;x_0)}{1+{k_0}^2 G^0_{d \rightarrow 1}(x_2,s;x_2)G^0_2(x_2,s;x_2)},
\end{equation}
where $G^0_2(x,s;x_0)=\left < x \left|[s-{\cal L}_2+k_r]^{-1}\right| x_0 \right>$ corresponds to the propagation of the particle on the second bridge potential put initially at $x_0$, in the absence of any coupling. So one can use the same procedure repeatedly to derive an analytical expression for the Greens function $G^{0}_{d \rightarrow a}(x,s;x_0)$ corresponding to the electron transfer from donor potential to acceptor potential through 'N' bridge potentials. 
\begin{equation}
G^{0}_{d \rightarrow a}(x,s;x_0)=G^0_{d \rightarrow N}(x,s;x_0) - \frac{{k_0}^2 G^0_{d \rightarrow N}(x,s;x_N)G^0_a(x_N,s;x_N)G^0_{d \rightarrow N}(x_N,s;x_0)}{1+{k_0}^2 G^0_{d \rightarrow N}(x_N,s;x_N)G^0_a(x_N,s;x_N)},
\end{equation}
where $G^0_a(x,s;x_0)=\left < x \left|[s-{\cal L}_a+k_r]^{-1}\right| x_0 \right>$ corresponds to the propagation of the particle on the acceptor potential put initially at $x_0$, in the absence of any coupling. All the potentials involved in this calculation are coupled to its nearest neighbouring potentials via Dirac delta functions (at $x_n$s) – so that we can use our two-potential result to analyse this problem. Using this Green's function one can calculate the correspondin ${\cal P}_{d \rightarrow a}(x,s)$ explicitely using Eq.(5). Here we are interested to know the survival probability at the donor potential 
\begin{equation}
P_{d \rightarrow a}(t) = \int_{-\infty}^{\infty} dx P_d(x,t).
\end{equation}
It is possible to evaluate Laplace Transform  ${\cal P}_d(s)$ of $P_d(t)$ directly. ${\cal P}_d (s)$ is defined in terms of ${\cal P}(x,s)$ by the following equation,
\begin{equation}
{\cal P}_{d \rightarrow a}(s) = \int_{-\infty}^{\infty} dx {\cal P}_d(x,s).
\end{equation}
Hence, we get
\begin{equation}
{\cal P}_{d \rightarrow a}(s)=\left(1-\left[1+k_0^2 G^0_{d \rightarrow N}(x,s;x_N)G^0_a(x_N,s;x_N)\right]^{-1}k_0^2 G^0_a(x_N,s;x_N)\int^{\infty}_{-\infty}dx_0 G^0_{d \rightarrow N}(x_N,s;x_0)P^0_d(x_0)\right)/(s+k_r).
\end{equation}
The average and long time rate constants can be found from ${\cal P}_{d \rightarrow a}(s)$ \cite{Kls2}. It is clear that, 
\begin{equation}
k^{-1}_{1}={\cal P}_{d \rightarrow a}(0).
\end{equation}
Hence,
\begin{equation}
k^{-1}_{1}=\left(1-\left[1+k_0^2 G^0_{d \rightarrow N}(x,0;x_N)G^0_a(x_N,0;x_N)\right]^{-1}k_0^2 G^0_a(x_N,0;x_N)\int^{\infty}_{-\infty}dx_0 G^0_{d \rightarrow N}(x_N,0;x_0)P^0_{d}(x_0)\right)/k_r.
\end{equation}
 On the other hand $k_{L}= - ($ pole of $\left[1-k_0^2 G^0_{d \rightarrow N}(x,s;x_N)G^0_a(x_N,s;x_N)(s+k_r)\right]^{-1})$, closest to the origin, on the negative $s$ - axis. The expression that we have obtained for ${\cal P}_{d \rightarrow a}(s)$, $k_I$ and $k_L$ are quite general and are valid for any set of potentials. In the following, we apply our method to the case of one dimensional semi-infinite periodic potential system. So we consider the electron transfer from $1$-st potential to $(n+1)$-th potential (where $n < N$), so there will be $(n+1)$ potential energy curves and $n$ coupling points (Dirac deltas). Now first we assume that all the potentials are equivalent and the distance between any two potential minima is the same, so the distance between any two neighbouring coupling points
will be the same. Let $G^0_n(x,0;x_0)$ denote the Green's function for the uncoupled motion of the electron on the $n$-th potential and $G^u_n(x,0;x_0)$ is defined by  $G^0_n(x,s;x_0)=\left < x \left|[s-{\cal L}_n+k_r]^{-1}\right| x_0 \right>$. Let $G^0_{1 \rightarrow n-1}(x,0;x_0)$ denote the Green's function for the coupled motion of electron on the first $n-1$ potentials. So the coupled motion of the electron on the first $n$ potentials can be analysed using our method. The Green's function for the coupled motion of electron on the first $n$ potential is given by
\begin{equation}
G^{0}_{1 \rightarrow n}(x,s;x_0)=G^0_{n}(x,s;x_0) - \frac{{k_0}^2 G^0_{n}(x,s;x_{n-1})G^0_{1\rightarrow n-1}(x_{n-1},s;x_{n-1})G^0_{n}(x_{n-1},s;x_0)}{1+{k_0}^2 G^0_{n}(x_{n-1},s;x_{n-1})G^0_{1 \rightarrow n-1} (x_{n-1},s;x_{n-1})}.
\end{equation}
Now, we consider the coupled motion of electron on the first $(n+1)$ potentials in the full system {\it i.e.} a total of $N+2$ potentials, where the $n+1$-th potential couples with $n$-th potential at $x_n$. For very large value of $n$ (semi-infinite limit) using the symmetry argument one can easily
prove that
\begin{equation}
G^{0}_{1 \rightarrow n}(x_n,s;x_n)=G^{0}_{1 \rightarrow n-1}(x_{n-1},s;x_{n-1}).
\end{equation}
So that, we get
\begin{equation}
G^{0}_{1 \rightarrow n}(x,s;x_0)=G^0_{n}(x,s;x_0) - \frac{{k_0}^2 G^0_{n}(x,s;x_{n-1})G^{0}_{1 \rightarrow n}(x_n,s;x_n) G^0_{n}(x_{n-1},s;x_0)}{1+{k_0}^2 G^0_{n}(x_{n-1},s;x_{n-1})G^{0}_{1 \rightarrow n}(x_n,s;x_n)}.
\end{equation}
If we put $x=x_n$ and $x_0=x_n$, in the above equation we get
\begin{equation}
G^{0}_{1 \rightarrow n}(x_n,s;x_n)=G^0_{n}(x_n,s;x_n) - \frac{{k_0}^2 G^0_{n}(x_n,s;x_{n-1})G^{0}_{1 \rightarrow n}(x_n,s;x_n) G^0_{n}(x_{n-1},s;x_n)}{1+{k_0}^2 G^0_{n}(x_{n-1},s;x_{n-1})G^{0}_{1 \rightarrow n}(x_n,s;x_n)}.
\end{equation}
The above equation can be simplified to
\begin{equation}
A(s) G^{0}_{1 \rightarrow n}(x_n,s;x_n)^2+ B(s) G^{0}_{1 \rightarrow n}(x_n,s;x_n) + C(s) = 0,
\end{equation}
where $A(s)= {k_0}^2 G^0_{n}(x_{n-1},s;x_{n-1})$, $B(s)= 1-{k_0}^2 G^0_{n}(x_n,s;x_n) G^0_{n}(x_{n-1},s;x_{n-1})+{k_0}^2 G^0_{n}(x_n,s;x_{n-1})G^0_{n}(x_{n-1},s;x_n)$ and $C (s) = - G^0_{n}(x_n,s;x_n)$. Now we solve the above equation for $G^{0}_{1 \rightarrow n}(x_n,s;x_n)$.
\begin{equation}
G^{0}_{1 \rightarrow n}(x_n,s;x_n)=\frac{- B(s) \pm \sqrt{B(s)^2 - 4 A(s) C(s)}}{2 A(s)} = f(s)
\end{equation}
In the above equation, $G^{0}_{1 \rightarrow n}(x_n,s;x_n)$ is expressed in terms of Green's function of an uncoupled potential. This when substituted back into Eq.(26), we get
\begin{equation}
G^{0}_{1 \rightarrow n}(x,s;x_0)=G^0_{n}(x,s;x_0) - \frac{{k_0}^2 f(s) G^0_{n}(x,s;x_{n-1}) G^0_{n}(x_{n-1},s;x_0)}{1+{k_0}^2 f(s) G^0_{n}(x_{n-1},s;x_{n-1})}.
\end{equation}
In the following, we apply our method for the case, where all potentials are assumed to be the parabolic potential, where we take 
\begin{equation}
V_n(x)= \frac{1}{2} B x^2 
\end{equation}
In this case, one may solve the following equation 
\begin{equation}
\left(s - {\cal L}_n + k_r \right) G^0_n(x,s;x_0)= \delta (x-x_0) 
\end{equation}
Using standard method \cite{Hilbert} to obtain.
\begin{equation}
G^0_n(x,s;x_0)=F(z,s;z_0)/(s+k_r)
\end{equation}
with
\begin{equation}
F(z,s;z_0)= D_\nu(-z_<)D_\nu(z_>)e^{(z_0^2-z^2)/4}\Gamma(1-\nu)[B/(2 \pi A)]^{1/2} 
\end{equation}
In the above, we have used a new variable $z$ defined by $z = x(A/B)^{1/2}$  and $z_j = x_j(A/B)^{1/2}$, $\nu  = —s/B$  and $\Gamma(\nu)$ is the gamma function.  $z_{<}= min(z, z_0)$ and $z_{>}= max(z, z_0)$. $D_{\nu}$ are parabolic cylinder functions. To get an understanding of the behavior of $k_I$ and $k_L$, we assume the initial distribution $P^0_e(x_0)$ is represented by $\delta(x-x_0)$. Then, we get 
\begin{equation}
{k_I}^{-1}= {k_r}^{-1}\left(1 - \frac{k_0^2 f(0)F(z_c,0;z_0)}{k_r+k_0^2 f(0)F(z_c,0;z_c)}     \right)
\end{equation}
It is important to mention that $k_I$ is dependent on the initial position $x_0$ (i.e., $z_0$), $k_r$ and $f(0)$. 
Again
\begin{equation}
k_L= k_r - [ values \; of \; s \; for \; which \;\; s+ k_0^2 f(s)F(z_c,s;z_c)=0]
\end{equation}
Again, $k_L$ is independent of $x_0$ and depends on $k_r$ and $f(s)$. So the long-term rate constant $k_L$ is determined by the
value of s, which satisfy $s+ k_0^2 f(s) F(z_c,s;z_c)=0$. We write this equation as an equation for $\nu (= -s /B)$
\begin{equation}
\nu = k_0^2 f(s) D_\nu(-z_c)D_\nu(z_c)\Gamma(1-\nu)/\sqrt{2 \pi AB}
\end{equation}
For integer values of $\nu$, $D_\nu(z)=2^{-\nu/2}e^{-z^2/4}H_{\nu}(z/\sqrt{2})$, $H_{\nu}$ are Hermite polynomials. $\Gamma(1-\nu)$ has poles at $\nu = 1,2, . . . .$. A graphical understanding using all these information shows that there is one value of $\nu \in [n,n+1]$ which satisfies Eq. (25). Our interest is in $\nu \in [0, 1]$, as $k_L =B \nu$ for $k_r=0$. If $k_0^2 f(s) /(2 \pi A B) \ll 1$, or $z_c \gg 1$ then $\nu \ll 1$ and one can arrive
\begin{equation}
\nu = k_0^2 f(s') D_0(-z_c)D_0(z_c)/\sqrt{2 \pi A B}
\end{equation}
and hence 
\begin{equation}
k_L = \sqrt{B/(2 \pi A)} k_0^2 f(s') e^{-{z_s}^2/2}.
\end{equation}
The same method can also be applied to the case where $S$ is a nonlocal operator, and may be represented by  $S=\left|f\right>k_0\left <g \right|$, where $f$ and $g$ are arbitrary acceptable functions. Choosing both of them to be Gaussian will be an interesting improvement over the current model. $S$ can also be a linear combination of such operators.

\end{document}